\documentclass[final,onefignum,onetabnum]{siamart171218}

\usepackage{lipsum}
\usepackage{amsfonts}
\usepackage{graphicx}
\usepackage{epstopdf}
\usepackage{algorithmic}
\ifpdf
  \DeclareGraphicsExtensions{.eps,.pdf,.png,.jpg}
\else
  \DeclareGraphicsExtensions{.eps}
\fi
\usepackage{graphicx}
\usepackage{mathtools}
\usepackage{amsmath}
\usepackage{amssymb}
\usepackage{xspace}
\usepackage{cleveref}
\usepackage{subcaption}
\usepackage[space]{grffile}
\usepackage{printlen}

\usepackage[square,sort,comma,numbers]{natbib}

\newsiamremark{remark}{Remark}
\newsiamremark{hypothesis}{Hypothesis}
\crefname{hypothesis}{Hypothesis}{Hypotheses}
\newsiamthm{claim}{Claim}

\headers{Transport Facilitated by Rapid Binding to Elastic Tethers}{B. Fogelson and J. P. Keener}

\title{Transport Facilitated by Rapid Binding to Elastic Tethers\thanks{Submitted to the editors September 1, 2018.
\funding{This research has been supported in part by the National Science Foundation under grants DMS 1515130 and DMS RTG 1148230, and in part by the Mathematical Biosciences Institute and the National Science Foundation under grant DMS 1440386. Any opinions, findings, and conclusions or recommendations expressed in this material are those of the authors and do not necessarily reflect the views of the National Science Foundation.}}}

\author{
	Ben Fogelson\footnotemark[2] \footnotemark[3]%
	\and
	James P. Keener\footnotemark[2]%
}

\DeclareCaptionLabelFormat{subreflabel}{\bfseries(#2)}
\usepackage{amsopn}

\providecommand{\Kappa}{K}

\newcommand{\pu}{\ensuremath{p_u}\xspace}
\newcommand{\pb}{\ensuremath{p_b}\xspace}
\newcommand{\qu}{\ensuremath{q_u}\xspace}
\newcommand{\qb}{\ensuremath{q_b}\xspace}
\newcommand{\ru}{\ensuremath{r_u}\xspace}
\newcommand{\rb}{\ensuremath{r_b}\xspace}

\newcommand{\pv}{\ensuremath{p_v}\xspace}
\newcommand{\pf}{\ensuremath{p_f}\xspace}

\newcommand{\subtether}[1]{\ensuremath{#1_\text{tether}}\xspace}
\newcommand{\ptether}{\subtether{p}}
\newcommand{\Dtether}{\subtether{D}}
\newcommand{\ktether}{\subtether{k}}
\newcommand{\zetatether}{\subtether{\zeta}}

\newcommand{\puhat}{\ensuremath{\hat{\pu}}\xspace}
\newcommand{\pbhat}{\ensuremath{\hat{\pb}}\xspace}
\newcommand{\xhat}{\ensuremath{\hat{x}}\xspace}
\newcommand{\yhat}{\ensuremath{\hat{y}}\xspace}
\newcommand{\zhat}{\ensuremath{\hat{z}}\xspace}
\newcommand{\Zhat}{\ensuremath{\hat{Z}}\xspace}
\newcommand{\that}{\ensuremath{\hat{t}}\xspace}

\newcommand{\vhat}{\ensuremath{\hat{v}}\xspace}
\newcommand{\Fhat}{\ensuremath{\hat{F}}\xspace}

\newcommand{\pvhat}{\ensuremath{\hat{\pv}}\xspace}
\newcommand{\pfhat}{\ensuremath{\hat{\pf}}\xspace}

\newcommand{\kon}{\ensuremath{k^+}\xspace}
\newcommand{\koff}{\ensuremath{k^-}\xspace}

\newcommand{\konhat}{\ensuremath{\hat{\kon}}\xspace}
\newcommand{\koffhat}{\ensuremath{\hat{\koff}}\xspace}

\newcommand{\intinf}{ \int_{-\infty}^{\infty}}

\newcommand{\kBT}{k_B T}

\newcommand{\Lbar}{\ensuremath{\bar{L}}\xspace}
\newcommand{\tbar}{\ensuremath{\bar{t}}\xspace}

\newcommand{\vbar}{\ensuremath{\bar{v}}\xspace}
\newcommand{\Fbar}{\ensuremath{\bar{F}}\xspace}

\newcommand{\Fbells}{\ensuremath{F_\text{unbinding}}\xspace}

\newcommand{\timescale}[1]{\ensuremath{\tau_\text{#1}}\xspace}
\newcommand{\tbinding}{\timescale{binding}}
\newcommand{\trelaxation}{\timescale{relaxation}}
\newcommand{\tdiffusion}{\timescale{diffusion}}

\DeclareMathOperator{\erfc}{erfc}

\ifpdf
\hypersetup{
  pdftitle={Transport Facilitated by Rapid Binding to Elastic Tethers},
  pdfauthor={Ben Fogelson and James P. Keener}
}
\fi

\begin{document}

\maketitle

\makeatletter
\long\def\@makefntext#1{\parindent 1em\noindent\hbox to 1.8em{\hss$\m@th^{\@thefnmark}$}#1}%
\makeatother

\renewcommand{\thefootnote}{\fnsymbol{footnote}}

\footnotetext[2]{Department of Mathematics, University of Utah, Salt Lake City, UT 84112}
\footnotetext[3]{Corresponding author: \email{ben@math.utah.edu}, \mbox{\url{http://math.utah.edu/\~ben}}}

\makeatletter
\long\def\@makefntext#1{%
    \parindent .25in%
    \noindent
    \hbox to .25in{\hss\@makefnmark}#1}
\makeatother

\renewcommand{\thefootnote}{\arabic{footnote}}

\begin{abstract}
  Diffusion in cell biology is important and complicated. Diffusing particles must contend with a complex environment as they make their way through the cell. We analyze a particular type of complexity that arises when diffusing particles reversibly bind to elastically tethered binding partners. Using asymptotic analysis, we derive effective equations for the transport of both single and multiple particles in the presence of such elastic tethers. We show that for the case of linear elasticity and simple binding kinetics, the elastic tethers have a weak hindering effect on particle motion when only one particle is present, while, remarkably, strongly enhancing particle motion when multiple particles are present. We give a physical interpretation of this result that suggests a similar effect may be present in other biological settings.
\end{abstract}



\section{Introduction}

Diffusion is one of the fundamental spatial processes driving biological function. In many cases, it is the primary mechanism by which cells distribute or transport particles, both within single cells and throughout tissues. Diffusion-driven morphogen gradients, in conjunction with chemical reactions and active transport, are responsible for directional signaling in development \cite{Shvartsman2012,Brooks2017}; diffusion of signaling molecules from the cell membrane to the nucleus is important for gene regulation \cite{Lindsay2017,Isaacson2016}; and within the nucleus, crowding by chromatin has a profound effect on the diffusion of proteins to specific DNA binding sites \cite{Isaacson2011}. There is a wealth of other biological examples. A unifying feature of diffusion in cell biology is that it occurs in a complicated environment rife with heterogeneity, complex mechanical interactions, and chemical binding.

In this paper, we examine the motion of diffusing particles that bind to and unbind from other objects as they move. Classical examples of this include the buffered diffusion of calcium ions as they reversibly bind to buffering proteins, resulting in a change to calcium's effective diffusion coefficient\cite{Wagner1994,Keener2008}, and the facilitated diffusion of oxygen into muscle fibers, where the total inward flux of oxygen is dramatically amplified by the binding of oxygen and myoglobin \cite{Rubinow1977,Keener2008}. If the particle of interest and its binding partner are both free to diffuse, as is the case in both of these examples, binding can dramatically enhance particle transport. If, however, the binding partner is anchored to a fixed substrate, it is straightforward to show that in the quasi-steady-state binding limit the particle's effective diffusion coefficient always decreases.

There is an important intermediate case that, to our knowledge, has not been studied. This is the case where the binding partner is anchored to a surface by an elastic tether, so that the binding partner has some freedom to move but does not diffuse over the entire domain. This situation is biologically relevant. It occurs in the nuclear pore complex in the interaction between karyopherin chaperones and elastic FG nucleoporins \cite{Fogelson2018, Raveh2016, Schmidt2016, Lim2007, Kimura2017}, and in a similar manner in the ciliary pore complex \cite{Pedersen2016,Takao2016,Kee2013}. It is also relevant to the diffusion of signaling molecules near membrane-bound cognate receptors, where elasticity can come from the mechanical properties of the receptor or from the membrane itself.

The fundamental question is how the mechanical properties of the elastic tether influence the particle of interest's overall motion. In classical facilitated and buffered diffusion, this overall motion is controlled by an interplay between the reaction rates governing particle binding and the diffusion coefficients of both the particle and its binding partner. The elastic properties of a tethered binding partner seem likely to alter this interplay in a way that is hard to predict a priori. On the one hand, we might intuit that the freedom of bound tethers to fluctuate about their base might cause them to bind particles ``at a distance,'' and thereby exert an elastic force that accelerates particle motion. On the other hand, a particle bound to a tether will not be able to move very far without unbinding.

In this paper, we develop and analyze a minimal model for investigating the interactions between particle diffusion, particle-tether binding, and tether elasticity. We begin by studying the motion of a single particle diffusing in one dimension against a background density of elastically tethered binding sites. We start with a Fokker-Planck system describing both the position and the binding state of the particle and, using a quasi-steady-state reduction, derive a reduced Fokker-Planck equation for the effective motion of the particle in the limit of fast binding and unbinding. Our approach is similar to the quasi-steady-state analysis of molecular motors developed by Newby and Bressloff \cite{Newby2010} and extended to nonlinear reaction terms by Zmurchok et. al.  \cite{Zmurchok2017}. The key difference is that our analysis explicitly accounts for spatial variations in the position of the binding sites, which changes the dimensionality of our starting system of equations. Our analysis shows that the particle's effective motion is controlled by a set of moments related to the space-dependent binding and unbinding rate functions. In particular, we show that for the simplest physical case of linearly elastic tethers, the particle's effective diffusion coefficient always drops.

Next, we generalize to the case where multiple diffusing particles are present. To study this, we apply a slightly modified version of a model that we recently developed for nucleocytoplasmic transport \cite{Fogelson2018}. We show that, in contrast to the single particle case, in the limit where binding is fast and the tethers are short compared to the domain of interest, the effective diffusion coefficient always increases. This qualitative difference between single and multiple-particle motion is surprising, and in the final part of the paper we give a physical interpretation of these results.

\section{One-dimensional motion of a single particle}
\label{sec:single}

\begin{figure}
	\centering
	\includegraphics{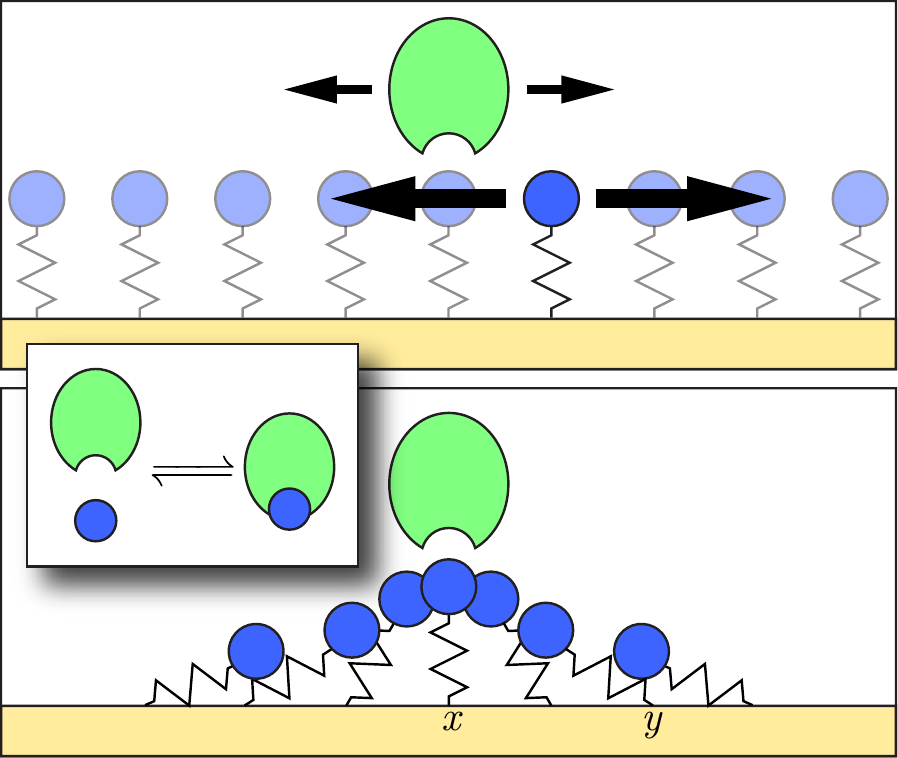}
	{
		\phantomsubcaption
		\label{fig:fastslow}
	}
	{
		\phantomsubcaption
		\label{fig:springsatdistance}
	}
	\caption{Schematic of single-particle diffusion mediated by elastic tethers. \subref*{fig:fastslow} A single particle (green) diffuses slowly in one dimension. A collection of elastically tethered binding sites (blue) diffuses much more rapidly than the particle. \subref*{fig:springsatdistance} Rapid diffusion of tethers means that a particle at position \(x\) has a non-zero probability per time of binding to a tether anchored at position \(y\).}
	\label{fig:diffusioncartoon}
\end{figure}

In this section, we derive an equation for the one-dimensional motion of a particle that undergoes diffusion while rapidly binding to and unbinding from a continuum of elastic tethers with constant density, as depicted schematically in \cref{fig:diffusioncartoon}. The starting point for this derivation is the set of differential Chapman-Kolmogorov equations
\begin{gather}
\frac{\partial \pu}{\partial t} = D \frac{\partial^2 \pu}{\partial x^2} - \pu \intinf \kon(y - x) \ dy + \intinf \koff(y - x) p_b(x, y, t) \ dy,
\label{eq:pudim}
\\
\frac{\partial \pb}{\partial t} = D \frac{\partial^2 \pb}{\partial x^2} - \frac{\partial}{\partial x} \biggl( \frac{k}{\zeta} (y - x) \pb \biggr) + \kon\bigl(y - x\bigr) \pu - \koff\bigl(y-x\bigr) \pb,
\label{eq:pbdim}
\end{gather}
where \(p_u(x, t)\) is the probability density that the particle is unbound and at location \(x\) at time \(t\) and \(p_b(x, y, t)\) is the probability density that the particle is at location \(x\) and is bound to a tether whose base is at \(y\). Together, \cref{eq:pudim,eq:pbdim} describe the evolution of these probabilities in terms of the particle's diffusion and drag coefficients \(D\) and \(\zeta\), and the tether's elastic spring constant \(k\). The functions \(\kon(z)\) and \(\koff(z)\) give the on and off rates for particle-tether binding, which we allow to be dependent on the spacing \(z\) between the particle and the tether base. We treat \kon and \koff as the overall reaction rates after accounting for the density of tether bases, so that we do not need to account for that density separately. The key assumption in our model is that an unbound tether diffuses much more rapidly than the particle, so that we do not need to track the position of individual tethers over time. Instead, we can represent the interaction between free tethers and particles with the binding rate \(\kon(z)\), which in general will depend on the equilibrium distribution of the tether head about its base. This is discussed for a linear spring in \cref{sec:diffusivemotionsingle}.

Our goal in this derivation is to describe the overall position of the particle, independent of its binding state. In other words, we seek an equation describing the evolution of the quantity
\begin{equation}
p(x, t) = \pu(x, t) + \intinf \pb(x, y, t) \ dy,
\end{equation}
which is the total probability density that the particle is at position \(x\) at time \(t\). We seek an approximate equation for \(p(x,t)\) that is valid in the limit of fast binding and unbinding.

\subsection{Nondimensionalization}

To approximate the particle's motion in the \linebreak limit of fast binding and unbinding, we nondimensionalize \cref{eq:pudim,eq:pbdim} in order to identify the relevant small parameter. This requires us to choose scales \Lbar and \tbar for length and time. Since the relevant lengthscale may vary depending on the specific physical application being modeled, we leave the lengthscale as \Lbar.

There are three important timescales: the binding timescale
\begin{equation}
	\tbinding = \biggl(\intinf \kon(z) \ dz\biggr)^{-1},
\end{equation}
the elastic relaxation timescale for a bound particle
\begin{equation}
	\trelaxation = \frac{\zeta}{k},
\end{equation}
and the diffusive timescale
\begin{equation}
	\tdiffusion = \frac{\Lbar^2}{D}.
\end{equation}
We expect that in most physical situations, binding dynamics are faster than both elastic and diffusive motion so that \(\tbinding \ll \trelaxation\) and \(\tbinding \ll \tdiffusion\). The relationship between \trelaxation and \tdiffusion is more subtle and is discussed below.

We choose to nondimensionalize time in units of the timescale \(\tbar = \tdiffusion\). Defining the functions \(\puhat(\xhat, \that)\) and \(\pbhat(\xhat, \yhat, \that)\), where \(\xhat = x/\Lbar\), \(\yhat = y/\Lbar\), and \(\zhat = z/\Lbar\), by
\begin{gather}
\pu(x, t) = \frac{1}{\Lbar}\puhat(x/\Lbar, t/\tbar)
\\
\pb(x, y, t) = \frac{1}{\Lbar^2}\pbhat(x/\Lbar, y/\Lbar, t/\tbar),
\end{gather}
the system \cref{eq:pudim,eq:pbdim} becomes
\begin{gather}
\frac{\partial \puhat}{\partial \that} = \frac{\partial^2 \puhat}{\partial \xhat^2} - \puhat \intinf \konhat(\yhat - \xhat) \ d\yhat + \intinf \koffhat(\yhat - \xhat) \pbhat(\xhat, \yhat, \that) \ d\yhat,
\\
\frac{\partial \pbhat}{\partial \that} = \frac{\partial^2 \pbhat}{\partial \xhat^2} - \frac{1}{\lambda} \frac{\partial}{\partial \xhat} \Bigl( (\yhat - \xhat) \pbhat \Bigr)+ \konhat(\yhat - \xhat) \puhat - \koffhat(\yhat - \xhat) \pbhat,
\end{gather}
where the nondimensionalized on and off rates \konhat and \koffhat are
\begin{gather}
\konhat(\zhat) = \Lbar \tbar \kon(\Lbar \zhat),
\\
\koffhat(\zhat) = \tbar \koff(\Lbar \zhat),
\end{gather}
and the non-dimensional parameter \(\lambda\) is
\begin{equation}
	\lambda = \frac{\trelaxation}{\tdiffusion}.
\end{equation}
The size of \(\lambda\) can be estimated as follows. We know from the Stokes-Einstein relation that \(D \zeta = \kBT\), so that \(\lambda\) can be rewritten as
\begin{equation}
	\lambda = \frac{\kBT}{k \Lbar^2},
\end{equation}
the ratio of thermal energy to the elastic energy stored in a spring stretched deformed by \Lbar. If this quantity is large, then the tether springs must be very weak on the lengthscale of interest. We do not expect weak springs to substantially alter the trajectory of the particle. Thus, the interesting dynamics will occur when \(\trelaxation < \tdiffusion\), and so \(\lambda < 1\).

We expect the binding rate functions \(\konhat\) and \(\koffhat\) to be large, and so we introduce the nondimensional parameter \(\varepsilon = \tbinding/\tdiffusion\) and define the rescaled rate functions \(\alpha\) and \(\beta\) by
\begin{gather}
\alpha(\zhat) = \varepsilon \konhat(\zhat),
\\
\beta(\zhat) = \varepsilon \koffhat(\zhat).
\end{gather}
Note that our definition of \(\varepsilon\) means that \(\intinf \alpha(\zhat) \ d\zhat = 1\), so we can write the nondimensional system of equations as
\begin{gather}
\frac{\partial \puhat}{\partial \that} = \frac{\partial^2 \puhat}{\partial \xhat^2} - \frac{1}{\varepsilon} \puhat + \frac{1}{\varepsilon} \intinf \beta(\yhat - \xhat) \pbhat(\xhat, \yhat, \that) \ d\yhat,
\\
\frac{\partial \pbhat}{\partial \that} = \frac{\partial^2 \pbhat}{\partial \xhat^2} - \frac{1}{\lambda} \frac{\partial}{\partial \xhat} \Bigl( (\yhat - \xhat) \pbhat \Bigr)+ \frac{1}{\varepsilon} \alpha (\yhat - \xhat) \puhat - \frac{1}{\varepsilon} \beta(\yhat - \xhat) \pbhat.
\end{gather}

In the following sections, we develop an approximate expression for the position of the particle in the limit  \(\varepsilon \ll 1\).%

\subsection{Splitting}

After dropping the hats over our nondimensional functions and variables, we seek to approximate the system of equations
\begin{gather}
\frac{\partial \pu}{\partial t} = \frac{\partial^2 \pu}{\partial x^2} - \frac{1}{\varepsilon} \pu + \frac{1}{\varepsilon} \intinf \beta(y - x) \pb(x, y, t) \ dy,
\label{eq:pu}
\\
\frac{\partial \pb}{\partial t} = \frac{\partial^2 \pb}{\partial x^2} - \frac{1}{\lambda}\frac{\partial}{\partial x} \Bigl( (y - x) \pb \Bigr)+ \frac{\alpha(y - x)}{\varepsilon} \pu - \frac{\beta(y-x)}{\varepsilon} \pb
\label{eq:pb}
\end{gather}
for \(\varepsilon \ll 1\).

In the limit \(\varepsilon \rightarrow 0\) we find the solution of \cref{eq:pu,eq:pb} into \(\pb = \frac{\alpha(y-x)}{\beta(y-x)} \pu\). Motivated by this, and using vector notation to describe the solution \([\pu, \pb]^T\), we introduce the reaction operator \(R\), defined by
\begin{equation}
R \left(
\left[
\begin{array}{c}
\pu
\\
\pb
\end{array}
\right]
\right)
=
\left[
\begin{array}{c}
-\pu + \intinf \beta(y-x) \pb \ dy
\\
\alpha(y - x) \pu - \beta(y-x) \pb
\end{array}
\right].
\label{eq:Rdef}
\end{equation}
This allows us to split the solution into a quasi-steady-state term \([\qu, \qb]^T\) in the nullspace of \(R\) and a remainder term \([\ru, \rb]^T\) in the range of R:
\begin{equation}
\left[
\begin{array}{c}
\pu
\\
\pb
\end{array}
\right]
=
\left[
\begin{array}{c}
\qu
\\
\qb
\end{array}
\right]
+
\left[
\begin{array}{c}
\ru
\\
\rb
\end{array}
\right].
\label{eq:splitting}
\end{equation}
The nullspace of \(R\) is spanned by \([1, \alpha(y-x)/\beta(y-x)]^T\). Defining the function \(\kappa(z) = \alpha(z)/\beta(z)\), the quasi-steady-state term can be written in terms of a scalar variable \(q(x, t)\) as
\begin{equation}
\left[
\begin{array}{c}
\qu(x, t)
\\
\qb(x, y, t)
\end{array}
\right]
=
\frac{1}{1 + \Kappa_0}
\left[
\begin{array}{c}
1
\\
\kappa(y - x)
\end{array}
\right]
q(x, t),
\label{eq:qvector}
\end{equation}
where \(\Kappa_0 = \intinf \kappa(z) \ dz\), and the prefactor \(\frac{1}{1+\Kappa_0}\) is a useful normalization.

For \([\ru, \rb]^T\) to be in the range of \(R\), we know from the Fredholm alternative that it must be orthogonal to the nullspace of the adjoint operator \(R^*\). It is straightforward to show that the adjoint operator is
\begin{equation}
R^* \left(
\left[
\begin{array}{c}
\pu
\\
\pb
\end{array}
\right]
\right)
=
\left[
\begin{array}{c}
-\pu + \intinf \alpha(y-x) \pb \ dy
\\
\beta \pu - \beta \pb
\end{array}
\right],
\end{equation}
where the appropriate inner product is
\begin{equation}
\left\langle
\left[\begin{array}{c}f_u\\f_b\end{array}\right],
\left[\begin{array}{c}g_u\\g_b\end{array}\right]
\right\rangle
= f_u g_u + \intinf f_b g_b \ dy.
\end{equation}
The nullspace of \(R^*\) is spanned by the vector \([1, 1]^T\), and so the remainder term must satisfy the orthogonality condition
\begin{equation}
\ru + \intinf \rb \ dy = 0.
\label{eq:orthogonalitycondition}
\end{equation}
This means we can write the remainder \([\ru, \rb]^T\) in terms of a single unknown function \(r(x, y, t)\), with \(\rb(x, y, t) = r(x, y, t)\) and \(\ru(x, t) = -\intinf r(x, y, t) \ dy\).

Substituting this splitting into \cref{eq:pu,eq:pb}, integrating \cref{eq:pb} over \(y\), and adding that to \cref{eq:pu}, we get the equation
\begin{equation}
\frac{\partial q}{\partial t} = \frac{\partial^2 q}{\partial x^2} - \frac{\Kappa_1}{\lambda(1 + \Kappa_0)} \frac{\partial q}{\partial x} - \frac{1}{\lambda} \frac{\partial}{\partial x} \biggl(\intinf (y - x) r(x, y, t) \ dy \biggr),
\label{eq:q}
\end{equation}
where \(\Kappa_1 = \intinf z \kappa(z) \ dz\).
Using this expression for \(\frac{\partial q}{\partial t}\) in \cref{eq:pb}, we also get
\begin{equation}
\begin{alignedat}{2}
\varepsilon \frac{\partial r}{\partial t} &\mathrlap{=-\alpha(y - x) \intinf r(x, y, t) \ dy - \beta(y-x) r(x, y, t)}
\\
&+ \varepsilon \Biggl[&&\frac{\partial^2 r}{\partial x^2} - \frac{1}{\lambda} \frac{\partial}{\partial x}\Bigl((y - x) r(x, y, t) \Bigr)
\\
&&& + \frac{\kappa(y-x)}{\lambda(1 + \Kappa_0)} \frac{\partial}{\partial x} \intinf (y - x) r(x, y, t) \ dy
\\
&&& + \frac{1}{1 + \Kappa_0} \frac{\partial^2}{\partial x^2} \Bigl( \kappa(y-x) q \Bigr) - \frac{1}{1 + \Kappa_0} \kappa(y - x) \frac{\partial^2 q}{\partial x^2}
\\
&&& + \frac{\Kappa_1}{\lambda (1 + \Kappa_0)^2} \kappa(y - x) \frac{\partial q}{\partial x} - \frac{1}{\lambda(1 + \Kappa_0)} \frac{\partial}{\partial x}\Bigl( (y - x) \kappa(y - x) q \Bigr) \Biggr].
\end{alignedat}
\label{eq:r}
\end{equation}
It is important to note that \cref{eq:q,eq:r} are exact. There is no approximation in going from \cref{eq:pu,eq:pb} to \cref{eq:q,eq:r}, only a change of variables from \pu and \pb to \(q\) and \(r\). Note also that the orthogonality condition \cref{eq:orthogonalitycondition} implies that
\begin{equation}
\pu(x, t) + \intinf \pb(x, y, t) \ dy = q(x, t),
\end{equation}
so that \(q\) is the total probability density that the particle is at position \(x\). This was the reason for the prefactor of \(\frac{1}{1 + \Kappa_0}\) in the definition of \(q\) in \cref{eq:qvector}.

\subsection{Approximation}

We seek an approximate equation for \(q\) that is valid when \(\varepsilon \ll 1\). To do this, we expand \(r\) as a power series in \(\varepsilon\):
\begin{equation}
r(x, y, t) = r_0(x, y, t) + \varepsilon r_1(x, y, t) + \mathcal O(\varepsilon^2),
\end{equation}
and solve for \(r_0\) and \(r_1\). From \cref{eq:r}, the zeroth order equation is
\begin{equation}
0 = -\alpha(y - x) \intinf r_0(x, y, t) \ dy - \beta(y - x) r_0(x, y, t).
\label{eq:r0}
\end{equation}
Dividing \cref{eq:r0} by \(\beta(y-x)\) and integrating gives
\begin{equation}
(1 + \Kappa_0) \intinf r_0(x, y, t) \ dy = 0.
\end{equation}
Substituting this back into \cref{eq:r0}, we get \(r_0(x, y, t) = 0\). This confirms that, as expected, the remainder term \(r\) is small when the reactions are fast. Substituting \(r_0 = 0\) into \cref{eq:r} gives the first order equation
\begin{equation}
\small
\begin{alignedat}{3}
0 &= -(1 + \Kappa_0)\Bigl(r_1 + \kappa(y-x) \intinf r_1(x, y, t) \ dy\Bigr)
+ \frac{1}{\beta(y - x)} \frac{\partial^2}{\partial x^2} \Bigl(\kappa(y - x) q \Bigr)
\\
&- \frac{\kappa(y - x)}{\beta(y - x)} \frac{\partial^2 q}{\partial x^2}
- \frac{1}{\lambda \beta(y-x)} \frac{\partial}{\partial x} \Bigr((y-x) \kappa(y-x) q\Bigr)
+ \frac{\Kappa_1}{\lambda(1 + \Kappa_0)} \frac{\kappa(y-x)}{\beta(y-x)} \frac{\partial q}{\partial x}.
\end{alignedat}
\label{eq:r1}
\end{equation}
We can integrate this equation over \(y\) to solve for \(\intinf r_1\). Noting that all the integrals share a similar form, we write
\begin{equation}
\phi_{i, j, k}(z) = \frac{z^i}{\beta(z)} \frac{\partial^j}{\partial z^j}\Bigl(z^k \kappa(z) \Bigr)
\end{equation}
for integers \(i\), \(j\), and \(k\), and let \(\Phi_{i, j, k} = \intinf \phi_{i, j, k}(z) \ dz\).

In terms of \(\Phi\), the integral of \(r_1\) is
\begin{equation}
\begin{alignedat}{1}
\intinf r_1(x, y, t) \ dy &= \frac{1}{\lambda (1 + \Kappa_0)^2} \biggl[\Bigl(\lambda \Phi_{0,2,0} + \Phi_{0,1,1} \Bigr) q(x, t)
\\
 &-\Bigl(2 \lambda \Phi_{0,1,0} + \Phi_{1,0,0} - \frac{\Kappa_1}{1 + \Kappa_0} \Phi_{0,0,0} \Bigr) \frac{\partial q}{\partial x}\biggr].
 \end{alignedat}
\end{equation}
Substituting this back into \cref{eq:r1} lets us solve for \(r_1\). Finally, substituting for \(r_1\) in \cref{eq:q}, we get a single differential equation for \(q\):
\begin{equation}
\small
\begin{alignedat}{2}
\frac{\partial q}{\partial t} &= \frac{\partial^2 q}{\partial x^2} - \frac{\Kappa_1}{\lambda(1 + \Kappa_0)} \frac{\partial q}{\partial x}
\\
&+ \frac{\varepsilon}{\lambda^2(1 + \Kappa_0)}\Biggl[
\biggl(
\frac{\Kappa_1}{1 + \Kappa_0}\bigl(\Phi_{0,1,1} + \lambda \Phi_{0,2,0}\bigr) - \bigl(\Phi_{1,1,1} + \lambda \Phi_{1,2,0}\bigr)
\biggr) \frac{\partial q}{\partial x}
\\
&+ \biggl(
\Phi_{1,0,1} + 2 \lambda \Phi_{1,1,0} - 2 \frac{\Kappa_1}{1 + \Kappa_0} \Bigl( \lambda \Phi_{0,1,0} + \Phi_{1,0,0}\Bigr) + \biggl(\frac{\Kappa_1}{1 + \Kappa_0}\biggr)^2
\Phi_{0,0,0} \biggr) \frac{\partial^2 q}{\partial x^2}
\Biggr]
\\
&+ \mathcal O(\varepsilon^2).
\end{alignedat}
\label{eq:qfinal}
\end{equation}
\Cref{eq:qfinal} is our desired result. This equation describes the evolution of the particle position to order \(\varepsilon\) purely in terms of \(q\) and integrals of the binding on and off rate functions. In the following sections, we analyze \cref{eq:qfinal} in some important special cases.

\subsection{Diffusive motion of a single particle}
\label{sec:diffusivemotionsingle}

\Cref{eq:qfinal} simplifies substantially when the binding and unbinding rates are spatially symmetric. When this is true, both \(\alpha(z)\) and \(\beta(z)\) are even functions of \(z\). This causes many of the integral terms in \cref{eq:qfinal} to evaluate to zero. Specifically,
\begin{equation}
\Kappa_1 = \intinf z \kappa(z) \ dz = \intinf z \frac{\alpha(z)}{\beta(z)} \ dz = 0,
\end{equation}
and
\begin{equation}
\Phi_{i,j,k} = \intinf \frac{z^i}{\beta(z)} \frac{\partial^j}{\partial z^j} \Bigl( z^k \kappa(z) \Bigr) \ dz = 0
\end{equation}
for \(i + j + k\) odd. Then, to order \(\varepsilon\), the equation for \(q\) becomes
\begin{equation}
\frac{\partial q}{\partial t} = \frac{\partial^2 q}{\partial x^2} + \frac{\varepsilon}{\lambda^2(1 + \Kappa_0)} \Bigl( \Phi_{1,0,1} + 2 \lambda \Phi_{1,1,0} \Bigr) \frac{\partial^2 q}{\partial x^2}.
\label{eq:qeven}
\end{equation}

Thus, when binding and unbinding are spatially symmetric, the net motion of the particle is purely diffusive. Moreover, fast binding to and unbinding from elastic tethers is capable of increasing or decreasing the particle's effective diffusion coefficient, depending on the sign of \(\Phi_{1,0,1} + 2 \lambda \Phi_{1,1,0}\), which in turn depends on the specific choices of the reaction rate functions \(\alpha\) and \(\beta\) as well as the ratio \(\lambda = \trelaxation/\tdiffusion\).

\subsubsection{Binding rate for a linear elastic spring} The simplest binding rates occur when the elastic tether is a simple linear spring and when the diffusion coefficient of a free tether is much larger than that of the particle. In this case, the dimensional steady-state probability \(\ptether(x; y)\) that the tether's free end is at position \(x\) with base fixed at \(y\) satisfies the differential equation
\begin{equation}
0 = \Dtether \frac{\partial^2 \ptether}{\partial x^2} - \frac{\ktether}{\zetatether}\frac{\partial}{\partial x} \Bigl( (y - x) \ptether \Bigr),
\end{equation}
and so \ptether is Gaussian:
\begin{equation}
\ptether(x; y) = \frac{1}{\sqrt{2 \pi}} \sqrt{\frac{\ktether}{\Dtether \zetatether}} e^{-\frac{1}{2} \frac{\ktether}{\Dtether \zetatether}(y - x)^2}.
\end{equation}
Here \Dtether, \zetatether, and \ktether are the diffusion, drag, and spring coefficients of the free tether. The Stokes-Einstein relation tells us that diffusion and drag are related by \(\Dtether = \frac{\kBT}{\zetatether}\), so
\begin{equation}
\ptether(x;y) = \frac{1}{\sqrt{2 \pi}}\sqrt{\frac{\ktether}{\kBT}} e^{-\frac{1}{2} \frac{\ktether}{\kBT} (y - x)^2}.
\end{equation}

We return to \ptether in a moment, but first we need to consider the basic kinetics of particle-tether binding. If the reaction simply occurred in a well-mixed solution of particles and tether binding sites (without the elastic anchors), the system would evolve according to mass-action kinetics as
\begin{equation}
	\frac{d c_\text{complex}}{d t} = k_\text{on} c_\text{particle} c_\text{tether} - k_\text{off} c_\text{complex},
\end{equation}
for linear densities of particles, tether binding sites, and particle-tether complexes \(c_\text{particle}\), \(c_\text{tether}\), and \(c_\text{complex}\), and rate constants \(k_\text{on}\) and \(k_\text{off}\). We can use the rate constant \(k_\text{on}\) along with \(\ptether\) to compute the separation-dependent binding rate function \kon.

Let the linear density of tether bases anchored to the domain be \(\rho\). For a small length \(dy\) the quantity \(\rho \, \ptether(x; y) \, dy\) can be interpreted as the expected density of tether free ends at \(x\) with bases in the interval \((y, y + dy)\). Since \(\pu\) is the probability density that there is a free particle at \(x\), the quantity \(k_\text{on}\, \rho \, \ptether(x; y) \, dy \, \pu(x, t)\) is the rate of increase of the probability density that there is a particle-tether complex with base in \((y, y + dy)\) at position \(x\). By definition, that is the rate of increase of the quantity \(\pb(x, y, t) \, dy\). Recalling \cref{eq:pbdim}, our original Fokker-Planck equation for the evolution of \pb, the kinetic rate constant \(k_\text{on}\) is thus related to the rate function \(\kon\) according to
\begin{equation}
	\kon(y - x) = k_\text{on}\, \rho \, \ptether(x; y) = k_\text{on}\, \rho \frac{1}{\sqrt{2 \pi}}\sqrt{\frac{\ktether}{\kBT}} e^{-\frac{1}{2} \frac{\ktether}{\kBT} (y - x)^2}.
	\label{eq:kongaussian}
\end{equation}
In nondimensional variables, this gives us the binding rate function
\begin{equation}
\alpha(y - x) = \sqrt{\frac{\ktether}{2 \pi k\lambda}} e^{-\frac{\ktether}{2k\lambda} (y-x)^2}.
\label{eq:alphagaussian}
\end{equation}
Note that \(\rho\) and \kon do not appear in \cref{eq:alphagaussian}. This is because \(\alpha\) was defined by  \(\alpha(\cdot) = \varepsilon \konhat(\cdot) = \frac{\tbinding}{\tdiffusion} \konhat(\cdot)\), and \(\tbinding = \left( \intinf \kon(z) \ dz \right)^{-1} = 1/(\rho k_\text{on})\).
The quantity \(\ktether/k\) is the ratio of the free tether spring coefficient to the bound tether spring coefficient. These two spring coefficients could be different due to conformational changes that occur upon binding.

\subsubsection{Diffusion mediated by linearly-elastic binding}
Taking the unbinding rate to be constant, we have the dimensional expression
\begin{equation}
	\koff(z) = k_\text{off},
\end{equation}
which is equivalent to the nondimensional expression
\begin{equation}
\beta(z) = \beta_0,
\label{eq:betaconstant}
\end{equation}
where \(\beta_0 = \tbinding k_\text{off} = \frac{k_\text{off}}{\rho k_\text{on}}\).

The evolution equation \cref{eq:qeven} for \(q\) simplifies to
\begin{equation}
\frac{\partial q}{\partial t} = \frac{\partial^2 q}{\partial x^2}  + \varepsilon \frac{\nu - 2}{\beta_0 (1 + \beta_0)\lambda} \frac{\partial^2 q}{\partial x^2},
\label{eq:qgaussian}
\end{equation}
where \(\nu = k/\ktether\) is the ratio of bound to unbound tether spring coefficients. \Cref{eq:qgaussian} predicts that for elastic tethers to increase the diffusion coefficient of a single particle, the inequality \(k > 2 \ktether\) must be satisfied, meaning that the tether must be at least twice as stiff when bound to a particle than when unbound. It is easy to envision how, in a biological setting, this could be the case: binding could induce a conformational change that stiffens the tether. In such a case, we expect that either the binding or unbinding step would require an input of chemical energy in the form of ATP. If, on the other hand, there is no conformational change due to particle-tether binding (so \(k = \ktether\) and therefore \(\nu = 1\)), \cref{eq:qgaussian} predicts that binding will hinder the particle's effective motion.

\begin{figure}
\centering
	\includegraphics{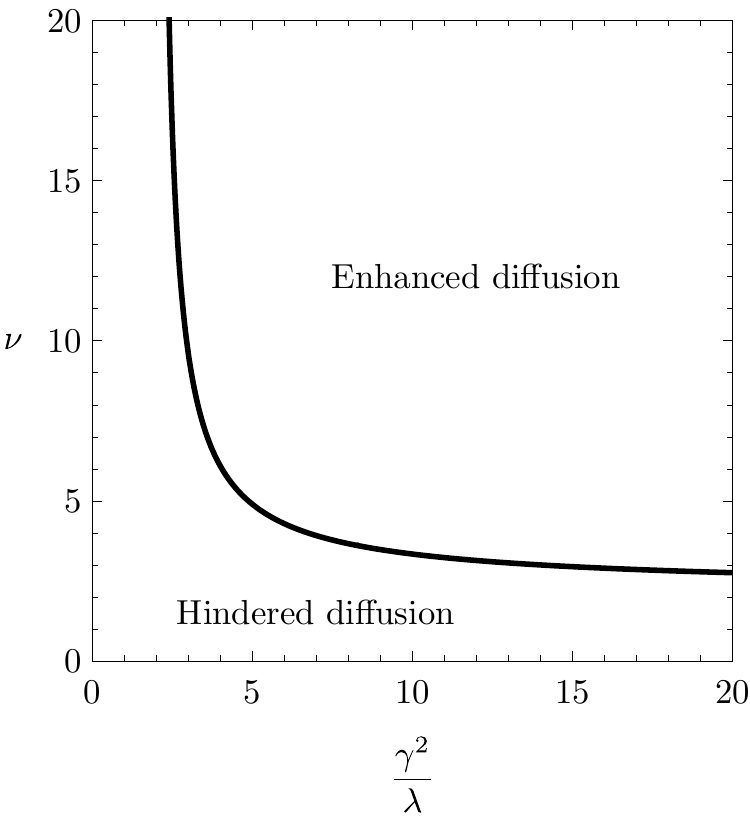}
	\caption{Parameter regimes for enhanced and hindered diffusion with slip bond unbinding given by \cref{eq:bells}.}
	\label{fig:zeroContour}
\end{figure}

Another reasonable choice for \(\koff(z)\) would be a classic slip bond with force-dependent unbinding \cite{Bell1978}:
\begin{equation}
	\koff(z) = k_\text{off} e^{\bigl(k |z|/\Fbells \bigr)},
	\label{eq:bellsdim}
\end{equation}
for some characteristic unbinding force \Fbells. In nondimensional units, \cref{eq:bellsdim} becomes
\begin{equation}
	\beta(z) = \beta_0 e^{\bigl( |z|/\gamma \bigr)},
	\label{eq:bells}
\end{equation}
where \(\beta_0 = k_\text{off} \tbinding\) and \(\gamma = \Fbells/(k \Lbar)\). With this unbinding rate function, the evolution equation for \(q\) becomes
\begin{equation}
	\scalebox{.98}{\(\displaystyle
	\frac{\partial q}{\partial t} = \frac{\partial^2 q}{\partial x^2} + \epsilon
	\left(
		\frac
			{
				2 \sqrt{2} (1 - \nu) \nu  \sqrt{\frac{\lambda  \nu }{\gamma ^2}} + \sqrt{\pi } \nu  e^{\frac{2 \lambda  \nu }{\gamma ^2}} \left(\frac{4 \lambda  (\nu -1) \nu }{\gamma ^2}+\nu -2\right) \erfc\left(\sqrt{\frac{2 \lambda  \nu }{\gamma^2 }}\right)
			}
			{
				\sqrt{\pi } \beta_0 \lambda  \nu \left(\beta_0+e^{\frac{\lambda  \nu }{2 \gamma ^2}} \erfc\left(\sqrt{\frac{\lambda  \nu }{2 \gamma^2 }}\right)\right)
			}
	\right) \frac{\partial^2 q}{\partial x^2}.
	\)}
	\label{eq:qbells}
\end{equation}
While this equation is clearly more complicated than when \(\beta(z) = \beta_0\), the qualitative predictions are similar. Note that the numerator of the correction term in \cref{eq:qbells} depends only on \(\nu\) and the quantity \(\gamma^2/\lambda\). Using this, we plot the parameter regimes for which \cref{eq:qbells} predicts enhanced and hindered diffusion in \cref{fig:zeroContour}. We can interpret \(\gamma^2/\lambda\) by rewriting it as \(\bigl(\Fbells^2/k\bigr)\big/\kBT\). The numerator here is the work it takes to stretch the tether by length \(\Fbells/k\), which is the characteristic lengthscale at which the bond begins to slip. Thus, as \cref{fig:zeroContour} shows, the larger the unbinding lengthscale the easier it is for diffusion to be enhanced. Indeed, in the limit as \(\gamma \rightarrow \infty\), \cref{eq:qbells} reduces to \cref{eq:qgaussian}, for which diffusion is enhanced when \(\nu > 2\). Additionally, when \(\nu = 1\), diffusion is hindered regardless of the value of \(\gamma^2/\lambda\). This is easily verified algebraically by substituting \(\nu = 1\) into \cref{eq:qbells}.

The main conclusions of this section are first, that for general spatially-dependent binding and unbinding rates, it is possible for a particle's overall motion to be enhanced by binding to a population of elastic tethers; and second, that in the simplest case where the tether mechanical properties are fixed and binding is governed by the linear elasticity of the tethers, the particle's diffusion is always reduced. A natural question, which we address in the following section, is whether there is some way to enhance particle motion even in the simple case.

\section{Motion of multiple particles}
\label{sec:multiple}

In \cite{Fogelson2018}, we proposed a model for enhanced transport through the nuclear pore that relied on competition between two distinct species of particles for binding to a population of elastic tethers. Enhanced nuclear transport in that case depended on a set of biochemically motivated boundary conditions as well as the interaction between two distinct species of moving particles.

Here, we derive a version of that model but with only one chemical species and with a scaling consistent with our analysis in \cref{sec:single}. We show that even when the elastic tethers are simple linear springs, intraspecies competition is sufficient to substantially enhance particle diffusion. Even more remarkably, we show that this enhancement effect is \(\mathcal O(1)\).

\subsection{Model derivation}

The derivation in this section follows the general approach we developed in \cite{Fogelson2018}. We proceed in three main steps: first, for a given spatial distribution \(v(x,t)\) of particles we derive an expression for the probability density that a tether anchored at \(y\) is bound to a particle at \(x\). Second, for an ensemble of tethers we estimate the expected force exerted on each particle. Third, we approximate this force to obtain a closed form expression for the flux of particles. From this flux we can write down a conservation law for the evolution of the particle distribution.

\subsubsection{Probability density for a tether to be bound}

Let \(v(x, t)\) be the concentration (number per unit length) of particles at position \(x\) and time \(t\), and consider a single elastic tether anchored at position \(y\). We can write an equation describing the probability of finding the tether bound to a particular particle:
\begin{equation}
\frac{\partial \pv}{\partial t} = \frac{1}{\rho}\kon\bigl(y-x\bigr) v(x, t) \pf(t; y) - \koff \bigl(y - x)\bigr) \pv(x, t; y).
\label{eq:pv}
\end{equation}
Here, \(\pv(x, t; y)\) is the probability density that the tether anchored at \(y\) is bound to a particle at \(x\) at time \(t\), and \(\pf(t; y)\) is the probability that the tether at position \(y\) is free. We have written the tether base coordinate \(y\) as a parameter rather than an independent variable to emphasize that we are considering a single tether fixed at \(y\). The binding rate includes a factor of \(\tfrac{1}{\rho}\) because our definition of \kon in \cref{sec:single} incorporates the effect of tether density, while in \cref{eq:pv} we are considering a single tether.

Note also that since our probabilities must sum to 1, we have
\begin{equation}
\pf(t; y) = 1 - \intinf \pv(x, t; y) \ dx.
\label{eq:pf}
\end{equation}

We can non-dimensionalize \cref{eq:pv,eq:pf} with the same scaling as in \cref{sec:single}:
\begin{gather}
	\epsilon \frac{\partial \pvhat}{\partial \that} = \frac{\vbar}{\rho} \alpha(\yhat - \xhat) \vhat(\xhat, \that) \pfhat(\that; \yhat) - \beta(\yhat - \xhat) \pvhat(\xhat, \that; \yhat),
	\\
	\pfhat(\that; \yhat) = 1 - \intinf \pvhat(\xhat, \that; \yhat) \ d\xhat.
\end{gather}
where \vbar is a concentration scale for the particles. Since \(\epsilon \ll 1\), we take \pv to be in quasi-steady-state:
\begin{equation}
	\pvhat(\xhat; \yhat) = \frac{\kappa(\yhat-\xhat) \vhat(\xhat, \that)}{\frac{\rho}{\vbar} + \intinf \kappa(\yhat - \xhat) \vhat(\xhat, \that) \ d\xhat}.
	\label{eq:pvsol}
\end{equation}

\Cref{eq:pvsol} can be interpreted as the fraction of time that an isolated tether anchored at \(\yhat\) spends bound to a particle at position \(\xhat\). When more than one tether is present, we expect there to be correlations between the binding states of different tethers. We ignore those correlations, and treat \cref{eq:pvsol} as a mean-field expression for the fraction of time that a tether at \(\yhat\) spends bound to a particle at \(\xhat\) even when other tethers are present.

\subsubsection{Expected force}

This approximation lets us estimate the average force on each particle due to tether binding. For a constant linear density of tethers \(\rho\), the total dimensional force on all particles at position \(x\) is
\begin{equation}
F(x, t) = \rho \intinf k(y-x) \pv(x; y) \ dy. %
\label{eq:forceintegral}
\end{equation}
For problems on this physical scale, the natural measure of energy is \(\kBT\), and so the natural scale for force is \(\Fbar = \kBT/\Lbar\). With this scaling, the nondimensional expression for the force is
\begin{equation}
	\Fhat(\xhat, \that) = \frac{\Lbar \rho}{\lambda} \intinf \frac{(\yhat - \xhat) \kappa(\yhat - \xhat)\vhat(\xhat, \that)}
	{\frac{\rho}{\vbar} + \intinf \kappa(\zhat) \vhat(\zhat, \that) \ d\zhat} \ d\yhat.
\end{equation}
With the expressions for \(\alpha(\cdot)\) and \(\beta(\cdot)\) from \cref{eq:alphagaussian,eq:betaconstant}, this becomes
\begin{equation}
	\Fhat(\xhat, \that) = \frac{\Lbar \rho}{\lambda}
	\intinf \frac{\frac{1}{\beta_0\sqrt{2\pi\lambda}} (\yhat - \xhat) \exp\bigl(-(\yhat-\xhat)^2/2\lambda \bigr) \vhat(\xhat,\that)}
	{\frac{\rho}{\vbar}+ \frac{1}{\beta_0\sqrt{2\pi\lambda}}\intinf \exp\bigl(-(\yhat-\zhat)^2/2\lambda\bigr) \vhat(\zhat, \that) \ d\zhat} \ d\yhat.
	\label{eq:fhat}
\end{equation}
The width of the Gaussians in \cref{eq:fhat} is governed by the parameter \(\lambda\), which we recall is \(\trelaxation/\tdiffusion\), the ratio of the elastic relaxation time to the time it takes for a free particle to have diffused a mean-squared distance of \Lbar. We have already argued that the physically interesting case is \(\lambda < 1\). Now we assume that \(\lambda \ll 1\), and approximate the integrals in \cref{eq:fhat} for small \(\lambda\).

Letting \(\Zhat = (\yhat-\xhat)/\sqrt{\lambda}\), we can rewrite the inner integral in \cref{eq:fhat}:
\begin{equation}
	\begin{alignedat}{2}
		\intinf \exp\bigl(-(\yhat-\zhat)^2/2\lambda\bigr) \vhat(\zhat, \that) \ d\zhat
		&=
		\intinf \sqrt{\lambda} \exp\bigl( -\Zhat^2/2 \bigr) \vhat(\yhat + \sqrt{\lambda} \Zhat, \that)\ d\Zhat
		\\
		&\approx \sqrt{2\pi\lambda} \vhat(\yhat, \that) + \mathcal O\bigl(\lambda^{3/2}\bigr).
	\end{alignedat}
	\label{eq:integralapproximation}
\end{equation}
Using a similar expansion for the outer integral, we get the expected force
\begin{equation}
	\Fhat(\xhat, \that) \approx -\Lbar \rho \frac{\vhat(\xhat,\that) \vhat_{\xhat}(\xhat, \that)}{\bigl(\beta_0 \frac{\rho}{\vbar} + \vhat(\xhat,\that) \bigr)^2} + \mathcal O(\lambda).
	\label{eq:fhatapproximated}
\end{equation}
We use this approximate expression for the force to compute the expected flux of particles.

\subsubsection{\texorpdfstring{Flux of \(\pmb v\)}{Flux of v}}

The quantity \(F(x)\) is the expected total force on all particles at position \(x\). This force will cause the particles to move. Recalling that \(\zeta\) is the particle drag coefficient, the expected flux of particles due to this elastic force is \(\frac{1}{\zeta} F(x)\). The total flux of particles is a combination of this elastic flux and normal Fickian diffusion:
\begin{equation}
	J(x, t) = -D v_x(x, t) + \frac{1}{\zeta} F(x, t).
\end{equation}
This leads to the conservation law
\begin{equation}
	\frac{\partial v}{\partial t} = D\frac{\partial}{\partial x} \biggl( -v_x + \frac{F(x, t)}{\kBT}\biggr).
\end{equation}
In non-dimensional variables, this becomes
\begin{equation}
	\frac{\partial \vhat}{\partial \that} = \frac{\partial}{\partial \xhat} \biggl(-\vhat_{\xhat} + \frac{\Fhat(\xhat, \that)}{\Lbar \vbar} \biggr),
\end{equation}
and using \cref{eq:fhatapproximated} for the force \Fhat, the conservation law becomes
\begin{equation}
	\frac{\partial \vhat}{\partial \that} = -\frac{\partial}{\partial \xhat} \left(\vhat_{\xhat} + \frac{\rho}{\vbar} \frac{\vhat \vhat_{\xhat}}{\Bigl(\beta_0 \frac{\rho}{\vbar} + \vhat \Bigr)^2}\right)
\end{equation}
Choosing the concentration scale \(\vbar = \frac{k_\text{off}}{k_\text{on}}\), and recalling that \(\beta_0 = \frac{k_\text{off}}{\rho k_\text{on}}\), we arrive at the equation
\begin{equation}
	\frac{\partial \vhat}{\partial \that} = - \frac{\partial}{\partial \xhat}\left( \vhat_{\xhat} + \frac{1}{\beta_0} \frac{\vhat \vhat_{\xhat}}{(1 + \vhat)^2} \right).
	\label{eq:vhat}
\end{equation}

\subsection{Enhanced diffusion}

\Cref{eq:vhat} can be interpreted as a nonlinear diffusion equation with the concentration-dependent diffusion coefficient \(1 + \frac{1}{\beta_0} \vhat/(1 + \vhat)^2\). This predicts that when \(\epsilon \ll 1\) and \(\lambda \ll 1\), diffusion is always enhanced by the competitive interaction between multiple particles and a density of elastic tethers. Interestingly, the form of the diffusion coefficient suggests an optimal particle concentration for enhanced diffusion. The diffusion coefficient grows like \(1 + \vhat/\beta_0\) for \(\vhat \ll 1\), reaches a maximum of \(1 + \tfrac{1}{4\beta_0}\) when \(\vhat = 1\), and decays back to 1 as \(\vhat \rightarrow \infty\).

This prediction of enhanced diffusion is in sharp contrast to the hindered diffusion in the single-particle case, where we found a diffusion coefficient of \(1 - \varepsilon \frac{1}{\beta_0(1+\beta_0)}\). This contrast is striking for two reasons. First, there is the simple fact that the same collection of elastic tethers slows down the motion of a single particle while speeding up the motion of an ensemble. Second, the magnitudes of the two effects are different: the hindrance in the single-particle case is only \(\mathcal O(\varepsilon)\), while the enhancement in the multiple-particle case is \(\mathcal O(1)\).

\subsection{Physical interpretation}

At first glance, it seems counter-intuitive that the effect of tether binding in the single and multiple particle cases would be different in both sign and order of magnitude. We can understand the physics driving this phenomenon by careful considering the force driving particle motion in each case.

\begin{figure}
	\centering
	{
		\phantomsubcaption
		\label{fig:cartoonA}
	}
	{
		\phantomsubcaption
		\label{fig:cartoonB}
	}
	\includegraphics[scale=1]{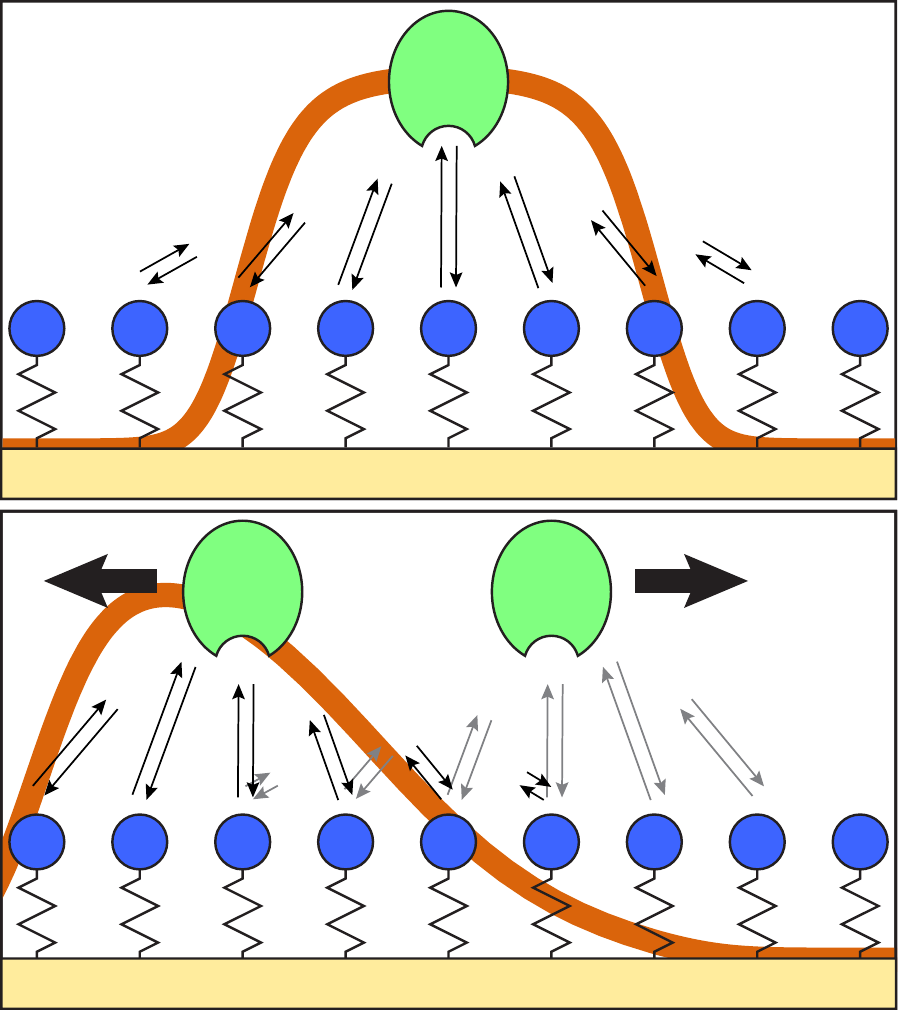}
	\caption{Schematic explanation of tether-facilitated diffusion with one and two particles. \subref*{fig:cartoonA} A single particle spends equal time bound to particles on either side of it, so it experiences a net elastic force. It is most likely to bind to tethers that are anchored directly underneath it (orange curve), which will tend to prevent the particle from drifting. \subref*{fig:cartoonB} When two particles compete to bind tethers, they spend a smaller fraction of time bound to the tethers in between them than to other tethers (orange curve). The net elastic force on each particle is non-zero, and causes the particles to repel one another (large arrows).}
	\label{fig:cartoon}
\end{figure}

Particles move due to a combination of Brownian motion driven by thermal fluctuations and directed motion driven by elastic tethers. When only one particle is present, as shown in \cref{fig:cartoonA}, symmetry means that the net elastic force on the particle is zero. Moreover, the particle spends most of its time bound to the tethers that are directly under it. Since the magnitude of the elastic tether force is proportional to the distance between the particle and the tether base, the particle feels little to no force from these tethers. Instead, these tethers anchor the particle and prevent it from drifting due to Brownian motion. While the particle does spend some amount of time bound to tethers that are anchored far away from the particle, the anchoring effect of binding to nearby tethers is dominant.

When multiple particles are present, competition for tether binding breaks the symmetry of the system. \Cref{fig:cartoonB} shows the case where two particles are near one another. The two particles can each bind to the tethers in between them, and so those tethers spend part of their time bound to each particle. This means that the particle on the left spends more overall time bound to tethers that are anchored to its left, and the particle on the right similarly spends more time bound to tethers to its right. Thus, the expected force on the left particle points to the left, and the average force on the right particle points to the right. For the two particles shown in \cref{fig:cartoonB}, this leads to an effective repulsive force between particles. For a collection of particles, that repulsion will push particles from regions of high concentration to regions of low concentration.

\section{Conclusion}

Understanding how particles move in the crowded, complex, and biochemically active environment of the cell is one of the major challenges of modern biology. We have developed a minimal model to explore one aspect of this challenge: how diffusion is influenced by frequent and reversible binding to elastic binding sites. Remarkably, our model predicts that the behavior of a single diffusing particle is dramatically different than the behavior of an ensemble. While the single particle's motion is hindered by binding, chemical competition within the ensemble generates an effective repulsion that drives particles down concentration gradients.

The key insight that explains this phenomenon is the observation that when binding sites are attached to flexible tethers, competition for binding occurs not just among particles at a single spatial location, but between particles at nearby locations. This nonlocal competition provides a mechanism by which particles ``sense'' the local concentration gradient: they are more likely to bind to an adjacent tether on the side with fewer neighbors, and so they are more likely to experience a force pulling them to that side.

While our model and analysis are limited to the one-dimensional case with linear springs for tethers, the underlying mechanism is quite general. Whenever elastic fluctuations introduce the possibility of binding at a distance, the system becomes sensitive to the local concentration gradient. This remains true even when the binding distance is quite short relative to lengthscales of interest, as was the case in our analysis.

In deriving our multiple particle model, we chose to ignore correlations between the binding states of individual tethers. It is unclear how significant these correlations are, and so it is unknown whether including them would change the qualitative predictions of our model. Future study and new analytical or computational techniques are necessary to answer this question.

To our knowledge, this quasi-steady-state analysis of a non-local representation of binding is new, and opens exciting avenues for future research. \Cref{eq:qfinal}, the reduced Fokker-Planck equation for single-particle motion, is valid for very general binding and unbinding rate functions, including the spatially anisotropic functions that might arise from complicated macromolecules such as molecular motors. We believe that our model framework, in both the single and multiple particle cases, will be a useful starting point for answering questions about transport in these more complicated settings.

\bibliographystyle{siamplain}
\bibliography{cleanlibrary}
\end{document}